\documentclass[letter,twocolumn]{jpsj3}

\usepackage{amsmath}
\usepackage{amssymb}
\usepackage{siunitx}
\usepackage{graphicx}

\title{Accelerating Extended Benders Decomposition with Quantum-Classical Hybrid Solver}

\author{Takuma Yoshihara$^{1,*}$ and Masayuki Ohzeki$^{1,2,3,4}$}

\inst{
$^{1}$Graduate School of Information Sciences, Tohoku University, Sendai 980-8564, Japan\\
$^{2}$Department of Physics, Science Tokyo, Meguro, Tokyo 152-8551, Japan\\
$^{3}$Research and Education Institute for Semiconductors and Informatics, Kumamoto University, Chuo, Kumamoto 860-0862, Japan\\
$^{4}$Sigma-i Co., Ltd., Minato, Tokyo 108-0075, Japan}

\abst{We propose a quantum–classical hybrid method for solving large-scale mixed-integer quadratic problems (MIQP). Although extended Benders decomposition is effective for MIQP, its master problem—which handles the integer and quadratic variables—often becomes a computational bottleneck. To address this challenge, we integrate the D-Wave CQM solver into the decomposition framework to solve the master problem directly. Our results show that this hybrid approach efficiently yields near-optimal solutions and, for certain problem instances, achieves exponential speedups over the leading commercial classical solver. These findings highlight a promising computational strategy for tackling complex mixed-integer optimization problems.}

\begin{document}
\maketitle

Optimization offers a powerful framework for decision-making by formulating complex problems as mathematical models. Among its many forms, Mixed-Integer Programming (MIP) is especially versatile, as it combines integer and continuous variables. This flexibility enables the modeling of real-world scenarios that require both discrete decisions and continuous adjustments.

In this study, we focus on Mixed-Integer Quadratic Programming (MIQP), where the objective function or constraints include quadratic terms. Quadratic formulations capture interactions between variables, making MIQP applicable to diverse and important domains. Notable examples include the unit commitment problem in power systems, which jointly optimizes generator schedules and power dispatch \cite{yang2016novel}, and cardinality-constrained portfolio optimization, which accounts for correlations between risk and return \cite{gao2013optimal}. These problems are highly challenging due to the combination of discrete and quadratic structures, demanding sophisticated solution methods. Classical approaches include branch-and-bound \cite{land1960automatic}, extended cutting-plane techniques \cite{westerlund1995extended}, and extended Benders decomposition (EBD) \cite{geoffrion1972generalized}.

EBD is a widely used and powerful technique for tackling MIQP. By decomposing the problem into a master problem and a subproblem, it iteratively converges to the optimal solution \cite{flippo1990note,lazimy1982mixed}. However, practical challenges remain: its convergence guarantees are limited, and solving the master problem often dominates the computational cost. When the master problem contains integer quadratic terms, iterative updates can even increase computational time, creating a severe bottleneck \cite{rahmaniani2017benders}.

This bottleneck is particularly pronounced in MIQP, where the master problem requires optimizing the quadratic form $x^T C x$ over discrete variables. Such problems quickly become intractable for classical solvers as the problem size grows. While recent efforts have attempted to accelerate master problem solving using machine learning \cite{lee2020ml_BGD} or quantum annealing \cite{zhao2022hybrid}, these approaches have been mainly limited to linear or simplified structures. The core difficulty in MIQP remains its quadratic interactions.

Quantum annealing (QA) \cite{kadowaki1998quantum}, a general-purpose solver for combinatorial optimization that exploits quantum tunneling, is naturally suited to quadratic objectives. Since the release of the first D-Wave device in 2011, QA has been applied to numerous domains, including finance \cite{orus2019finaceqa}, traffic optimization \cite{shikanai2025quadratic}, manufacturing scheduling \cite{haba2022travel,venturelli2015jobshopqa}, logistics \cite{ding2021logics}, materials discovery \cite{doi2023exploration}, and even procedural content generation for games \cite{ishikawa2023individual}. To extend its applicability, hybrid approaches that combine QA with conventional optimization methods have also been proposed \cite{hirama2023efficient,takabayashi2024hybrid}.

Building on these developments, we propose a new optimization method that leverages the Constrained Quadratic Model (CQM) solver developed by D-Wave Systems to solve the EBD master problem. The CQM solver is a hybrid framework that integrates quantum annealing with classical optimization, designed to efficiently produce high-quality solutions for NP-hard mixed-integer quadratic problems.

To evaluate the effectiveness of our approach, we conduct comparative experiments with existing methods. Specifically, we benchmark the proposed method against simulated annealing and a state-of-the-art classical MIP solver (Gurobi Optimizer). 
By comparing solution quality and computation time, we demonstrate that our hybrid method offers a promising and scalable strategy for solving large-scale MIQPs.

The master problem is MIQP, which is represented by the following objective function, Eq. (1), and constraint inequality (2).
\begin{align}
  \min_{x,y} \ \ &x^\top C x + h^\top y \\
  \text{s.t.} \ \ &Ax+Gy\leq b ,\\
  &x \in \{0,1\}^n , 
  y \in \mathbb{R}_{+}^m \notag.
\end{align}
where $C$ is a $n\times n$ symmetric matrix  and $h$ is a $m\times n$ nonnegative vector.  Here $A$ is a $m\times n$ and $G$ is a $m\times n$ constraint matrix, and $b$ is an $m$-dimensional constant vector. \par 
The first term $x^\top C x $ with discrete variables is difficult to solve with a classical solver, but we set it up this way because it is effective to solve it using QA. The continuous variable part $h^\top y$ and the constraints are linear and easy to solve with a linear solver.

In EBD, the solution to the master problem can be obtained iteratively and alternately using the solutions obtained for the upper and lower problems. First, assume that the solution $\overline{x}$ is obtained in the master problem, substitute the $\overline{x}$ into Eq. (1) and (2), and define the problem with the continuous variable $y$ as the lower-level problem. To further simplify the problem, we apply  Lagrange relaxation to the subproblem and, using the weak duality theorem, consider the dual problem using the dual variable $\lambda$ (Eq. (3), (4)).
\begin{align}
  \max_{\lambda} \ \ &(b-A\overline{x})^\top \lambda\\
  \text{s.t.} \ \ &G^\top \lambda \leq h, \\
  &\lambda \in \mathbb{R}_{+}^m \notag
\end{align}
If the lower dual problem can be solved, the solution $\lambda$ is the extreme point $u^k$. It is known from the weak duality theorem that $(b-A\overline{x})^\top u^k \leq h^\top y$. On the other hand, if there is no solution, instead of Eq. (4), we solve $G^\top \lambda \leq 0$ as a constraint and let its solution $\lambda$ be the extreme line $r^j$. 

Next, using the solution $u^k$ of the lower dual problem, consider the master problem consisting of an integer variable $x$ and a continuous variable $t$.
\begin{align}
  \min_{x,t} \ \ &x^\top C x + t \\
  \text{s.t.} \ \ &(b-Ax)^\top u^{k} \leq {t} \ \ \text{for} \ \ k \in \hat{K},\\
    \ \ &(b-Ax)^\top r^{j} \leq 0 \ \ \text{for} \ \ j \in \hat{J},\\
  &x\in \{0,1\}^n , t\in \mathbb{R} \notag.
\end{align}
An auxiliary variable $t$ is introduced in place of $h^\top y$, and the optimality constraint Eq. (6) is added to satisfy the weak duality theorem described above. 
Furthermore, by the weak duality theorem, the objective function value of the dual problem is known not to exceed that of the original problem, and it is known that $(b-A\overline{x})^\top u^k \leq h^\top y$ is valid.
If no extreme point is obtained, the dual subproblem obtains $ uk$ or the extreme line $ rj$; the constraints in Eqs. (6) and (7) corresponding to these values are not imposed.

Our algorithm EBD convergence occurs when the difference between the objective function value $t$ of the master problem and the objective function value $(b-Ax)^\top u^{k}$ of the dual subproblem falls below a certain threshold. 
In other words, this refers to the point at which the duality gap between the primal and dual problems becomes small. Therefore, it can be expressed as:
\begin{align}
    |t - (b-Ax)^\top u^{k} |\leq \epsilon
\end{align}
is obtained.
\par
Here, $\epsilon$ is a pre-set threshold serving as the criterion for judging EBD convergence. When the difference between $t$ and $(b-Ax)^\top u^{k}$ becomes less than or equal to $\epsilon$, it is judged that the optimal solution has been obtained with sufficient accuracy, and the iteration is terminated. In this study, $\epsilon$ is fixed at 0.5 , following the reference in previous research papers \cite{zhao2022hybrid}.\par
Thus, by repeatedly solving the master problem and the dual subproblem using EBD, convergence of the solution is achieved, ultimately yielding the desired solutions $x$ and $y$ for the main problem.\par

Equation (5) of the objective function of the master problem is NP-hard because it has a quadratic term in a discrete variable represented by $x^\top C x$, which accounts for most of the computational complexity of the EBD when the master problem is solved with a classical solver. Therefore, to solve the EBD efficiently, QA is used for the master problem. 
Quantum annealing effectively solves quadratic unconstrained binary optimization (QUBO) with binary variables. However, quantum annealing can only deal with discrete, binary variables and cannot contain constraints. The master problem has a continuous variable $t$ in Eq. (5), and Eq. (6) and (7) are inequality constraints, so the master problem must be converted to QUBO form.

The continuous variable $t$ in Eq. (5) is discretized using the binary variable $w$ as in Eq. (8).
\begin{align}
    t(w) &= \sum_{i=-\underline{m}}^{\overline{m}_{+}} 2^{i} w_{(i+\underline{m})} - \sum_{j=0}^{\overline{m}_{-}} 2^{j}w_{j+1+\underline{m} + \overline{m}_{+}} \\
    & w \in \{0,1\} ^{M} \notag
\end{align}
where $\overline{m}_{+}$ and $\overline{m}_{-}$ are the number of bits representing the positive and negative integer parts of the continuous variable $t(w)$,
$\underline{m}$ is the number of bits assigned to the positive fractional part. When the continuous variable $t$ is negative, it can be expressed by subtracting the negative value from the positive value of Eq. (9). Equation (9) for $t$, which consists only of the binary variable expression $w$, can be converted to QUBO form by substituting it into Eq. (5).

Next, the inequality constraint (6) is converted to an equality constraint by introducing a non-negative slack variable as a linear sum of binary $s_{kl}^ {K}$. Furthermore, by moving all terms of the equality constraint to one side and squaring them, we replace the QUBO form as in Eq. (10) and add them to Eq. (5).

\begin{align}
  P_k \left( \overline{t}(w) + (u^{k})^\top A x
  + \sum_{\,\ell=0}^{\overline{\ell}^{K}} 2^{\ell} s_{k\ell}^{\,K}
  - b^\top u^{k} \right)^{2},
  \intertext{where}
  \overline{\ell}^{K}
  = \left\lceil
      \log_{2}\!\left(
        b^\top u^{k}
        - \min_{w,x}\!\left( \overline{t}(w) + (u^{k})^\top A x \right)
      \right)
    \right\rceil \notag.
\end{align}

Here, $P_k$ is a positive constant indicating the magnitude of the constraint; if Eq. (6) is satisfied, a non-negative slack variable is found, and Eq. (10) is zero. In contrast, if Eq. (6) is not satisfied, Eq. (10) takes a positive value and is not employed in the problem of finding the minimum value as in Eq. (5). Equation (7) can be converted to QUBO format in the same way as above. It can be seen that Eq. (5) can be converted to QUBO form by introducing the binary variables $w$ and $s_{kl}^ {K}$ in Eq. (9).

In this study, we focus on scaling with respect to binary variables and constraints in the master problem. The number of continuous variables primarily affects the subproblem (a linear programming task), which can be solved efficiently by classical solvers and does not constitute the computational bottleneck addressed by our hybrid approach.


We conducted two sets of experiments. The first examined the convergence behavior and solution accuracy of the proposed EBD-based method. 
The second evaluated the computation time of different solvers applied to the master problem.  

We first investigated the convergence of our method using four solvers: Quantum Annealing (Advantage\_6.4, D-Wave Systems), Simulated Annealing (SASampler, OpenJij), the CQM hybrid solver (hybrid\_constrained\_quadratic\_model\_version1, D-Wave Systems), and the Gurobi Optimizer (v12.0.1). Note that the detailed algorithm of the CQM hybrid solver has not been made public.

\begin{figure}[tb]
    \centering
    \includegraphics[width=\linewidth]{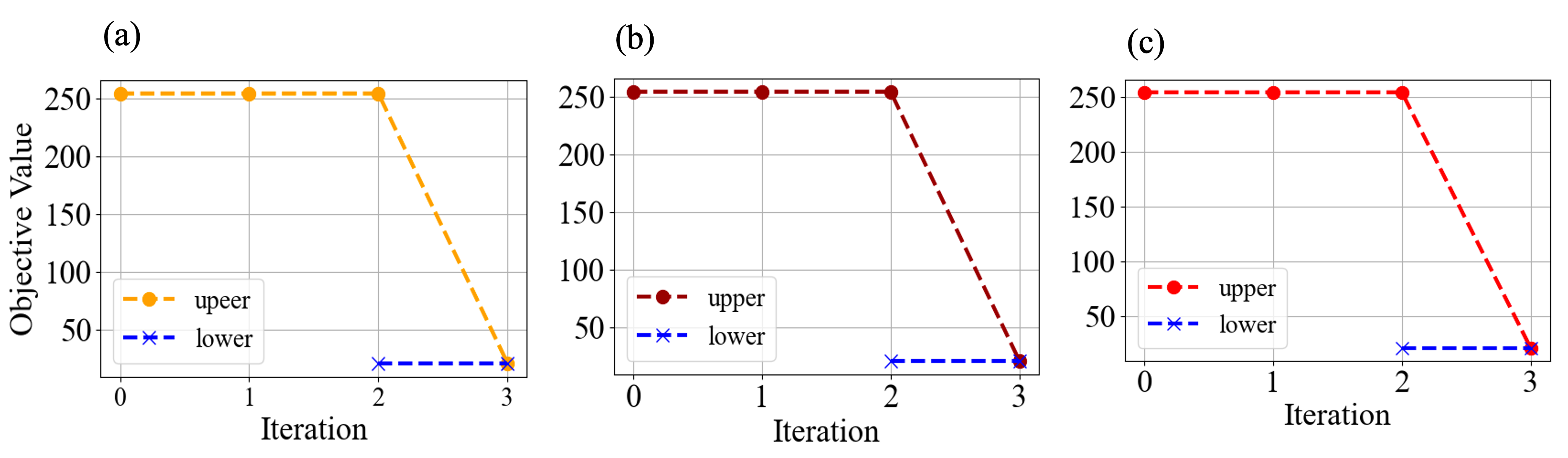}
    \caption{(Color online) Convergence processes of EBD using (a) Simulated Annealing, (b) Gurobi, and (c) CQM.}
    \label{fig:a}
\end{figure}

In this setting, the number of integer variables $x$ and continuous variables $y$ was fixed at five, with five constraints. The convergence results are shown in Fig.~\ref{fig:a}.  
For Simulated Annealing (SA), the parameters were set as $P_k = P_j = 0.55$, $\text{num\_reads} = 3000$, and $\text{num\_sweeps} = 3000$. Here, $\text{num\_reads}$ denotes the number of annealing trials used to generate candidate solutions, and $\text{num\_sweeps}$ specifies the number of full spin updates performed per trial.  

As shown in Fig.~\ref{fig:a}, SA, CQM, and Gurobi all follow nearly identical convergence trajectories, with final solutions in agreement. Thus, any of these solvers can ensure convergence of the proposed algorithm. In contrast, QA failed to converge, indicating that its current performance cannot provide the solution accuracy required for EBD.  

An additional strength of the proposed method lies in its ability to recover exact solutions. For MIQP problems with complicating variables, EBD provides an equivalent reformulation and guarantees global optimality under convexity and strong duality \cite{geoffrion1972generalized}. Our experiments confirmed that the solutions obtained by EBD with Gurobi matched those obtained by directly solving the same problem with Gurobi, thereby validating correctness.  

\begin{figure}[tb]
    \centering
    \includegraphics[width=\linewidth]{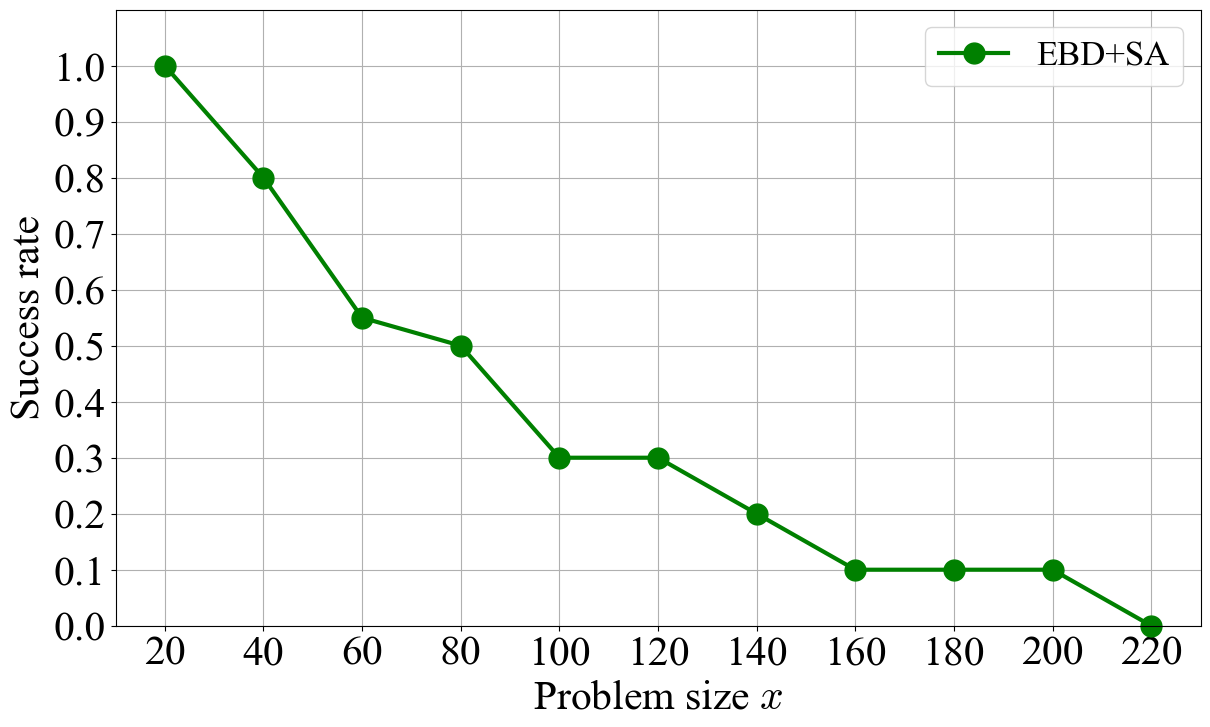}
    \caption{(Color online) Convergence success rate of SA with increasing problem size in EBD.}
    \label{fig:b}
\end{figure}

However, SA exhibited poor scalability. As the problem size increased, convergence could no longer be maintained. To examine this limitation, we tested 20 instances while varying problem size, fixing the number of constraints at five (increasing to ten would have required prohibitive time). This setup intentionally relaxed problem difficulty to assess whether SA could succeed under lenient conditions. The results are shown in Fig.~\ref{fig:b}.  

As shown in Fig.~\ref{fig:b}, SA converges reliably up to size 20, but its success probability decreases thereafter, and at size 220 convergence is essentially lost. This failure arises from the inherent limitations of the QUBO formulation and the inability of SA (and similarly QA) to maintain the precision EBD requires.  

\begin{figure}[tb]
    \centering
    \includegraphics[width=\linewidth]{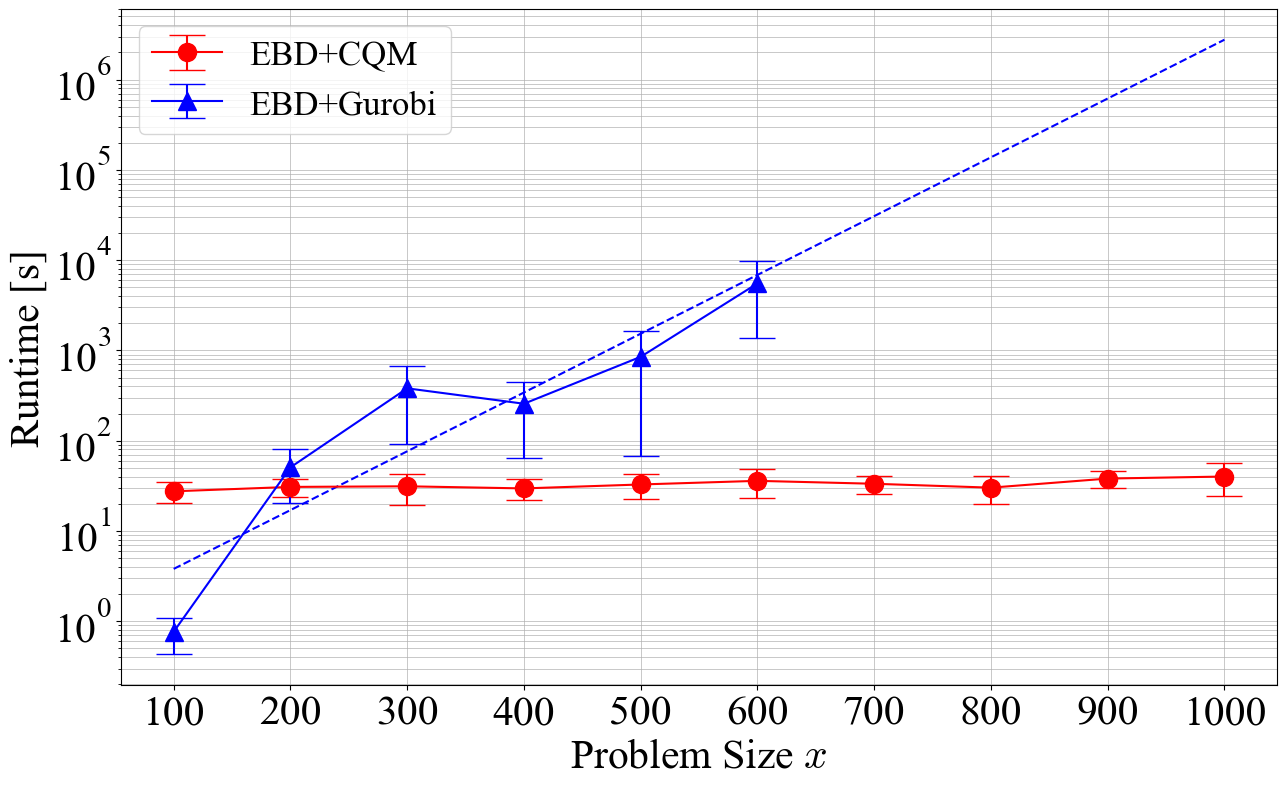}
    \caption{(Color online) Computational time for solving the EBD master problem using Gurobi and CQM as problem size increases from 100 to 1000. Gurobi treats values up to 600 as actual measurements and uses scaling plots for values beyond that.}
    \label{fig:c}
\end{figure}

Next, we compared computation times between CQM and Gurobi, the two solvers that consistently converged. Experiments were conducted by varying the number of integer variables $x$, while fixing the number of continuous variables $y$ at 10 and constraints at 10. Results are presented in Fig.~\ref{fig:c}.  

Figure~\ref{fig:c} shows that computation time grows rapidly with problem size for Gurobi, while the increase is significantly moderated for CQM. This demonstrates that when combined with CQM, the proposed method can solve the EBD master problem substantially faster, especially for large-scale MIQP instances.  
Furthermore, regarding the scaling behavior of the CQM solver, the results obtained in this study are considered acceptable based on prior research\cite{quinton2025quantum} , which suggests that CQM exhibits favorable scaling characteristics for quadratic binary problems compared to classical exact solvers.

\begin{figure}[tb]
    \centering
    \includegraphics[width=\linewidth]{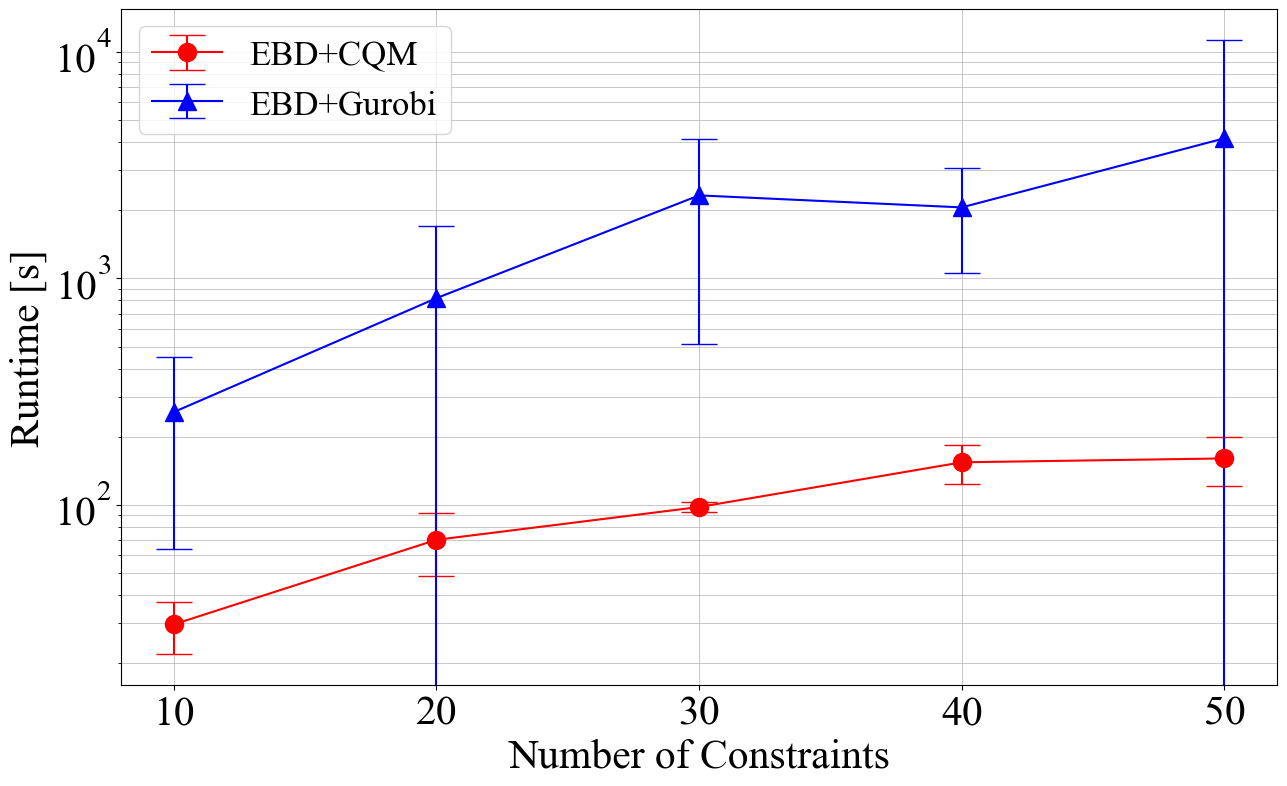}
    \caption{(Color online) Comparison of computation times for solving the master problem when fixing the binary variable to 400 and increasing the number of constraints.}
    \label{fig:d}
\end{figure} 

Figure~\ref{fig:d} shows the results of investigating how increasing the number of constraints affects computation time when the problem size for binary variables is fixed at 400. It demonstrates that CQM achieves faster computation speeds than Gurobi when the number of constraints is increased.

This study demonstrates that combining EBD with the CQM solver enables the generation of fast and exact solutions for MIQP. The performance was comparable to solving the problem directly with Gurobi, showing that, as the problem grows, the CQM solver can solve MIQP faster than Gurobi. In other words, the EBD+CQM approach provides a powerful means of tackling large-scale problems that have been difficult to address with conventional classical solvers.

This advantage makes the method particularly promising for real-world applications, since it allows natural scalability without performance degradation. By leveraging the flexibility of CQM in handling quadratic formulations together with the iterative computation strength of EBD, we anticipate that significantly larger problem instances can be tackled efficiently.

Future work will focus on developing this approach into a general-purpose framework for a wide range of mixed-integer problems. To achieve this, we will investigate the algorithm’s behavior under more complex continuous variables and an increasing number of constraints, in order to clarify the scalability limits of the proposed method. Defining these boundaries will allow us to determine which class of problems can benefit most from this framework.

In particular, we plan to evaluate its effectiveness on representative real-world MIQP problems such as power system optimization and cardinality-constrained portfolio optimization. Through these studies, we expect to establish EBD+CQM as a versatile and promising method for tackling mixed-integer problems in both academic research and practical applications.

We received financial supports by programs for bridging the gap between R\&D and IDeal society (Society 5.0) and Generating Economic and social value (BRIDGE) and Cross-ministerial Strategic Innovation Promotion Program (SIP) from the Cabinet Office (No. 23836436).

\bigskip
\noindent
$^*$yoshihara.takuma.p4@dc.tohoku.ac.jp

\bibliographystyle{jpsj}
\bibliography{sample}

\end{document}